\documentclass[11pt,a4paper]{article}
\usepackage[utf8]{inputenc}
\usepackage[T1]{fontenc}
\usepackage{lmodern}
\usepackage[margin=1in]{geometry}
\usepackage{amsmath,amssymb,amsthm}
\usepackage{booktabs}
\usepackage{graphicx}
\usepackage{hyperref}
\hypersetup{colorlinks=true,linkcolor=blue,citecolor=blue,urlcolor=blue}

\newtheorem{theorem}{Theorem}
\newtheorem{proposition}[theorem]{Proposition}
\newtheorem{corollary}[theorem]{Corollary}
\theoremstyle{remark}
\newtheorem{remark}[theorem]{Remark}

\newcommand{\eps}{\epsilon}
\newcommand{\Pml}{P^{\mathrm{ML}}_{\mathrm{fail}}}

\title{How wide must an eavesdropper be?\\Leakage of a toric-code logical bit to a strip\\and the Nishimori point}
\author{Konstantinos Karagiannis\thanks{Senior Director, Quantum Computing Services, Protiviti. Email: konstantinos.karagiannis@protiviti.com.}}
\date{\today}

\begin{document}
\maketitle

\begin{abstract}
Fault-tolerant quantum computers hand pieces of each logical qubit to separate decoder modules and shared readout lines. How much does a piece reveal? We prove that for any region and any Pauli noise, including correlated noise, an observer's optimal advantage in learning a logical bit is exactly $1-2\Pml$, where $\Pml$ is the failure probability of maximum-likelihood decoding confined to that region. Two lessons follow. Geometry, not noise, provides the protection: a region that misses a single column of the code learns exactly nothing at any noise level, while a region that spans the code is protected by noise only at impractical distances ($d\gtrsim350$ for one row at a $10^{-3}$ flip rate). And width is a poor guide to risk: under i.i.d.\ noise a two-row strip learns exactly as little as one row, yet below threshold an eavesdropper needs a width growing only as $\ln d$. Through the identity, the strip's operational threshold is the Nishimori point of the random-bond Ising model, and a Walsh--Hadamard transfer matrix up to width 16 locates it to within a few thousandths. On two IBM Heron processors ($d=3,5$), an emulated strip adversary shows the predicted jump at width three, and non-uniform qubit errors let a trained attacker profit from the second row that uniform noise makes useless. We turn these results into design rules for partitioned decoding and multiplexed readout.
\end{abstract}

\section{Introduction}

A logical qubit in a topological code is hidden from local observers: no measurement on a region smaller than the code distance reveals anything about it \cite{Kitaev2003,Dennis2002,BPT2010,BravyiTerhal2009}. In the language of quantum secret sharing, the code defines an access structure in which ``unauthorized'' regions learn nothing and ``authorized'' regions can reconstruct \cite{CleveGottesmanLo1999}. For the toric code without noise the access structure is purely topological: a region learns the logical $Z$ bit if and only if it supports a non-contractible cycle of the right homology class, and the information is then complete.

The noiseless access structure answers a binary question: does a region support a logical representative or not? Noise makes the problem richer. A region may support the logical operator yet be unable to read its value reliably, while a fairly narrow region may read it with high confidence by exploiting the redundancy inside its boundaries. What is missing is a quantitative description of this intermediate regime: not whether a region has access to the logical bit, but how reliably it can learn it.

The question is practical. Fault-tolerant machines decode in parallel, handing regions of the lattice to separate classical decoder modules \cite{Lin2024,Skoric2023,Tan2023}, and superconducting devices share readout lines whose crosstalk can reveal other qubits' outcomes to a co-located user \cite{Maurya2024,Mehra2024}. Either way, a party that should not learn a logical result can end up holding the readout of part of the code. How much does that part reveal, and how should a system be partitioned so that it reveals nothing?

The answers are not what intuition suggests. A region can hold every column of the code but one and learn exactly nothing about the logical bit, at any noise level. A region that spans the code is the opposite case: noise protects the bit only at code distances far beyond any planned device. At a physical flip rate of $10^{-3}$, a single row of qubits still holds half the maximum advantage at $d\approx350$, and three rows at $d\sim4\times10^4$. Below threshold, the width an eavesdropper needs grows only logarithmically with the code distance. And under uniform noise a second row adds exactly nothing to the first. On real hardware it does help, because the second row lets an attacker route around the worst qubits. All of this follows from one exact statement: what a region learns about the logical bit is precisely what a decoder confined to that region can recover. The statement matters for any setting where a code is used as a hiding device rather than only as a memory, and it sharpens what ``local indistinguishability'' means once decoherence is present, a subject that has recently been studied through global diagnostics such as coherent information and Markov length \cite{Bao2023,Fan2023,SangHsieh2024}.

We make four contributions.

\paragraph{1. Leakage is local optimal decoding.} For any region $A$ and any Pauli channel, including correlated ones, the trace-distance leakage $\eps_A$ equals $1-2\Pml(A)$, where the decoder is restricted to the stabilizers and logical representative supported in $A$ (Theorem~\ref{thm:identity}). We call this the \emph{regional leakage--decoding identity}: an observer's optimal advantage in distinguishing the two logical states from a region of a noisy code is fixed exactly by the performance of maximum-likelihood decoding confined to that region. The proof is elementary: both reduced states are diagonal in the joint eigenbasis of the supported stabilizer group, so distinguishing them is a classical problem. The theorem converts an information-theoretic question into a decoding question, with two immediate consequences: local noise can never create leakage, and the shape of $A$ matters only through the code it carries.

\paragraph{2. The strip: exact small-width results.} The simplest region that can leak is a strip of $w$ rows that wraps the torus. It carries a cylinder surface code of circumference $d$ whose distance against logical-flipping errors is $w$. Hence $1-\eps=\Theta(d\,p^{\lceil w/2\rceil})$ (Proposition~\ref{prop:smallp}), and a two-row strip learns no more than a single row (Corollary~\ref{cor:tworow}).

\paragraph{3. Width scaling and the threshold.} For bit-flip noise we compute $\eps(d,w,p)$ for $d\le 40$ and $w\le 8$, and the quenched and annealed rates for $w\le16$. The leakage decays as $e^{-d\varphi(w,p)}$; by the identity, the threshold of $\eps$ at fixed aspect ratio is the optimal decoding threshold, the Nishimori point $p_N=0.1094$ \cite{Nishimori1981,Honecker2001}, and a finite-size-scaling collapse of the quenched rate places the crossing at $p_c\approx0.105$--$0.108$ across fit choices, a few thousandths below $p_N$, which we attribute to finite-width corrections. This yields a three-regime answer to the question in the title. For partitioned decoding and shared readout lines, this becomes a set of design rules (Sec.~\ref{sec:design}): one missing column guarantees zero leakage, and noise alone does not.

\paragraph{4. Hardware.} We prepare encoded states of the rotated surface code on two IBM Heron processors in five runs over three days and emulate a strip adversary offline on the measured data (Sec.~\ref{sec:hw}); this tests regional readout leakage under real device noise, not a physical side-channel attack. The jump at $w=3$ is observed on every run. A second row, which adds nothing under uniform noise, helps a learned attacker on hardware, and independent but non-uniform flips with fitted per-qubit rates account for $90$--$99\%$ of that gain.

We emphasize what is not new. That subsystems of a code can be decoded, that logical failure rates combine an energetic cost set by the code distance with an entropic factor counting failure paths \cite{Beverland2019}, and that the optimal threshold of the toric code under bit-flip noise is the Nishimori point of the random-bond Ising model \cite{Dennis2002,WangHarringtonPreskill2003} are all known. Our contribution is the exact identity between region leakage and local decoding, its consequences for strip-shaped adversaries, and quantitative finite-size results validated by independent methods.

\section{Setting}

\subsection{Code, state and region}
Qubits sit on the $n=2d^2$ edges of a $d\times d$ square lattice on the torus. We write $h(i,j)$ for the horizontal edge from vertex $(i,j)$ to $(i,j+1)$ and $v(i,j)$ for the vertical edge from $(i,j)$ to $(i+1,j)$, indices mod $d$. Stars and plaquettes are
\begin{align}
A_{i,j}&=X_{h(i,j)}X_{h(i,j-1)}X_{v(i,j)}X_{v(i-1,j)},\\
B_{i,j}&=Z_{h(i,j)}Z_{h(i+1,j)}Z_{v(i,j)}Z_{v(i,j+1)}.
\end{align}
The encoded bit is the eigenvalue of $Z_0=\prod_j Z_{h(0,j)}$; the second logical qubit is fixed in the $+1$ eigenstate of $Z_1=\prod_i Z_{v(i,0)}$. The two states $|\psi_b\rangle$, $b\in\{0,1\}$, are stabilized by $\mathcal S_b=\langle A_{i,j},B_{i,j},(-1)^bZ_0,Z_1\rangle$.

For a region $A$ of edges, let $\rho_A(b)$ be the reduced state after an i.i.d.\ Pauli channel $\Phi=\bigotimes_q\Phi_q$ with $\Phi_q(\rho)=\sum_{P}p_P\,P\rho P$. The leakage is
\begin{equation}
\eps_A=\tfrac12\|\rho_A(0)-\rho_A(1)\|_1,
\end{equation}
and by Helstrom's theorem the optimal probability of guessing $b$ from $A$ is $(1+\eps_A)/2$ \cite{NielsenChuang2000}.

We consider bit-flip noise ($p_X=p$), dephasing ($p_Z=p$) and depolarizing noise ($p_X=p_Y=p_Z=p/3$). The \emph{flip probability} $q$ of an edge is the probability of an error that anticommutes with $Z$: $q=p$ for bit-flip and $q=2p/3$ for depolarizing.

\subsection{The strip}
The strip of width $w$ is
\begin{equation}
A_w=\{h(i,j):0\le i<w\}\cup\{v(i,j):0\le i<w-1\},\qquad |A_w|=(2w-1)d.
\end{equation}
It wraps the torus horizontally and therefore supports $Z_0$. It supports the $(w-1)d$ plaquettes $B_{i,j}$ with $0\le i\le w-2$ and the $(w-2)d$ stars $A_{i,j}$ with $1\le i\le w-2$, and nothing else from $\mathcal S_b$ other than products of these with $Z_0$. The strip with $w=1$ is the minimal row.

\subsection{Threat model}\label{sec:threat}
The adversary obtains the measurement outcomes of the data qubits in a region $A$ at the moment the logical qubit is read out, and nothing outside $A$. It knows the code, the region and the noise model, and has unbounded classical computing power; it cannot act on the quantum state. Two features of fault-tolerant architectures make this model concrete.

\emph{Partitioned classical decoding.} Real-time decoding at scale is parallelized by assigning separate decoder modules to overlapping spatial regions of the device \cite{Lin2024} and to windows in time \cite{Skoric2023,Tan2023}. Syndrome data alone reveal nothing: stabilizer outcomes are independent of the logical state, so a module that only ever sees syndromes learns nothing about it at any noise level. The exposure arises when outcomes whose product is a logical operator pass through a module, which happens in two routine situations: the destructive transversal readout of the data qubits that ends a logical measurement, and lattice-surgery merges, where the joint logical outcome is the product of the new stabilizers measured along the seam \cite{Horsman2012}. A module that is compromised, or whose input stream is observed, during such a readout sees exactly the outcomes in its region.

\emph{Readout side channels.} On superconducting devices, frequency-multiplexed readout shares a feedline among several qubits, and the resulting readout crosstalk lets a co-located user infer the measured states of other qubits on the line \cite{Maurya2024}; crosstalk-mediated attacks on shared hardware are an active concern more broadly \cite{Mehra2024}. An adversary with such a channel obtains noisy copies of the outcomes on the qubits it couples to.

When the noise acts on the readout as independent bit flips (bit-flip noise, or any channel whose $X$ and $Z$ components are independent), the optimal measurement on $A$ in Theorem~\ref{thm:identity} is itself a $Z$-basis readout of $A$ followed by classical decoding. The leakage $\eps_A$ is then exactly the advantage of this classical adversary, and side-channel inference errors simply add to the flip rate. Under correlated noise such as depolarizing noise, an adversary able to measure star operators inside $A$ can do slightly better (Sec.~\ref{sec:depol}), so the classical adversary's advantage is a lower bound on $\eps_A$. An adversary that can also measure stabilizers inside $A$ is covered by Theorem~\ref{thm:identity}, whose optimal measurement includes them; Sec.~\ref{sec:depol} quantifies what this adds under depolarizing noise.

\emph{Regional exposure.} In this model the exposure of a region is set not by how many physical qubits it contains but, through Theorem~\ref{thm:identity}, by whether it contains a logical representative and how well a decoder confined to it corrects that representative's errors. Qubit counts mislead in both directions: a region holding all but one column of the code learns nothing (Proposition~\ref{prop:column}), while a single row retains an advantage close to one at realistic distances and error rates (Table~\ref{tab:dhalf}).

\section{Leakage equals local maximum-likelihood decoding}

Let $G_A=\{g\in\mathcal S_0:\operatorname{supp}g\subseteq A\}$. Consider an arbitrary Pauli channel $\Phi(\rho)=\sum_E P(E)\,E\rho E^\dagger$, where $P$ is any probability distribution over $n$-qubit Pauli errors: correlated, non-identical and non-independent errors are all allowed. Because $EgE^\dagger=(-1)^{[E,g]}g$ for every Pauli error $E$, the channel maps each stabilizer element to $c(g)\,g$ with the average character $c(g)=\sum_EP(E)(-1)^{[E,g]}$. For independent single-qubit channels this factorizes as $c(g)=\prod_{q\in\operatorname{supp}g}c_q(g_q)$ with $c_q(P)=1-2\Pr[\text{error on }q\text{ anticommutes with }P]$. The stabilizer reduced-state formula \cite{Fattal2004} then gives
\begin{equation}
\rho_A(b)=2^{-|A|}\sum_{g\in G_A}c(g)\,(-1)^{b\,[g,X_0]}\,g .
\label{eq:rhoA}
\end{equation}
If $A$ supports a representative $L$ of $Z_0$, write $G_A=S_A\cup LS_A$, where $S_A$ (rank $r$) consists of the elements commuting with $X_0$.

\begin{theorem}[Regional leakage--decoding identity]\label{thm:identity}
For every region $A$ and every Pauli channel (in the general sense above),
\begin{equation}
\eps_A=1-2\Pml(A)=2^{-r}\sum_{\sigma\in\{0,1\}^r}\Big|\sum_{s\in S_A}c(Ls)\,(-1)^{\sigma\cdot s}\Big|,
\label{eq:identity}
\end{equation}
where $\Pml(A)$ is the failure probability of the optimal decoder that measures the generators of $S_A$ and the operator $L$, and guesses $b$ from the outcomes. If $A$ supports no representative of $Z_0$, $\eps_A=0$ at every noise strength.
\end{theorem}

\begin{proof}
All elements of $G_A$ commute, so both states in Eq.~\eqref{eq:rhoA} are diagonal in the joint eigenbasis of a generating set $\{g_1,\dots,g_r,L\}$. Each joint eigenspace, labelled by syndrome $\sigma$ and $L$-eigenvalue $z$, has dimension $2^{|A|-r-1}$, and the probability of outcome $(\sigma,z)$ in state $b$ is $2^{-r-1}\big[\sum_{s}c(s)\chi_\sigma(s)+(-1)^bz\sum_s c(Ls)\chi_\sigma(s)\big]$ with $\chi_\sigma(s)=(-1)^{\sigma\cdot s}$. The trace distance of two commuting states is the total-variation distance of these outcome distributions, which is the second expression in Eq.~\eqref{eq:identity}. The optimal guess of $b$ from $(\sigma,z)$ succeeds with probability $\tfrac12(1+\mathrm{TV})$, which is $1-\Pml$. If $A$ supports no representative of $Z_0$, no element of $G_A$ anticommutes with $X_0$ and the two states coincide.
\end{proof}

\begin{remark}
The vanishing of $\eps_A$ for regions without a logical representative holds for every Pauli channel, including correlated ones, and for product channels also follows from data processing. The content of Theorem~\ref{thm:identity} is the converse direction, and it is an equality rather than a bound: when $A$ supports the logical, the observer's optimal advantage is exactly what a decoder confined to $A$ achieves. The optimal measurement is also simple: measure the stabilizers supported in $A$ and the logical string, then decode classically. A quantity defined through reduced quantum states thus has an exact operational meaning as a classical decoding problem, and the correspondence does not require the Pauli errors to be independent.
\end{remark}

\paragraph{The whole logical qubit.} The same formula extends to an arbitrary logical state of the encoded qubit. Every element of $\mathcal S_0$ times a logical operator contributes to $\rho_A$ only if it is supported in $A$, so $\rho_A$ depends on the logical state only through the expectation values of the logical Paulis that have a representative in $A$. A strip that wraps the torus horizontally supports representatives of $Z_0$ but of neither $X_0$ nor $Y_0$, so for any two logical states $\tau,\tau'$,
\begin{equation}
\tfrac12\|\rho_A(\tau)-\rho_A(\tau')\|_1=\tfrac12\big|\langle Z_0\rangle_\tau-\langle Z_0\rangle_{\tau'}\big|\,\eps_A ,
\end{equation}
which is maximized by the two $Z_0$ eigenstates. In the language of secret sharing, a strip learns one Pauli component of the encoded qubit, attenuated by $\eps_A$, and nothing about the complementary ones; learning the qubit requires a region that also supports a vertical cycle.

For bit-flip noise the stars drop out: $c_q(X)=1$, so the sum over the star part of $S_A$ only enforces that their syndrome be trivial, and $\eps_A$ reduces to the classical problem of decoding the $Z$-type code on $A$ against i.i.d.\ flips with probability $q$. The same reduction holds for any channel with independent $X$ and $Z$ components. For depolarizing noise the stars carry information about $Y$ errors, which are correlated with flips; we return to this in Sec.~\ref{sec:depol}.

\section{Strips: exact small-width results}

Strip width gives a clean way to separate physical access from logical recovery. Widening a strip adds physical qubits and changes which stabilizers the observer holds, but neither guarantees a larger distinguishing advantage. The two-row strip is the sharpest example: under i.i.d.\ Pauli noise it holds more qubits than a single row and gives exactly the same advantage.

\begin{corollary}[Two rows learn no more than one]\label{cor:tworow}
For every i.i.d.\ Pauli channel with flip probability $q$ and every $d$,
\begin{equation}
\eps_{A_1}=\eps_{A_2}=(1-2q)^d .
\end{equation}
\end{corollary}
\begin{proof}
$A_1$ supports only $Z_0$, so $\eps_{A_1}=|c(Z_0)|=(1-2q)^d$ (the piling-up lemma \cite{Matsui1993}). $A_2$ supports the $d$ plaquettes $B_{0,j}$ and no stars, so only the flip pattern $e$ on the $3d$ edges matters. A flip on $h(0,j)$ and a flip on $h(1,j)$ violate exactly the same plaquette $B_{0,j}$ and are equally likely, but only the former flips the measured $Z_0$. Write $f_j=e_{h(0,j)}\oplus e_{h(1,j)}$. Conditioned on the syndrome, if any $f_j=1$ the parity of the row-0 flips is uniformly random; if all $f_j=0$ it is biased by $\prod_j[(1-q)^2-q^2]/[(1-q)^2+q^2]$. Averaging over syndromes, $\eps_{A_2}=\Pr[f=0]\cdot\big(\tfrac{1-2q}{(1-q)^2+q^2}\big)^d=(1-2q)^d$.
\end{proof}

The two-row strip has a single logical-flipping error of weight $2$ per column, $X_{h(0,j)}X_{h(1,j)}$, that commutes with every supported plaquette. More generally the strip $A_w$ is a surface code on a cylinder whose undetectable logical-flipping errors are vertical strings crossing all $w$ rows, so its distance against them is $w$.

\begin{proposition}[Small-$p$ scaling]\label{prop:smallp}
Let $A$ support a representative of $Z_0$, and let $\delta_A$ be the minimum weight of a flip pattern on $A$ that violates no $Z$-type element of $S_A$ and flips the measured logical, i.e.\ the distance of the code that $A$ carries against flips. Under i.i.d.\ bit-flip noise, $1-\eps_A=\Theta(p^{\lceil\delta_A/2\rceil})$ as $p\to0$. For the strip, $\delta_{A_w}=w$, so $1-\eps_{A_w}=\Theta(p^{\lceil w/2\rceil})$, with a prefactor proportional to $d$ for $d\gg w$.
\end{proposition}
\begin{proof}[Sketch]
By Theorem~\ref{thm:identity}, $1-\eps=2\Pml$. A code of distance $\delta_A$ decoded by maximum likelihood fails at leading order on configurations of $\lceil\delta_A/2\rceil$ flips on a minimum-weight undetectable logical (with ties contributing one half when $\delta_A$ is even). For the strip these are the vertical strings crossing all $w$ rows, of which there are $\Theta(d)$. The exponent is confirmed numerically by exact enumeration over all syndromes for $w=2,\dots,5$ (fitted exponents $0.98,1.98,2.04,3.00$; Appendix~\ref{app:checks}).
\end{proof}

\section{Width scaling and the threshold}\label{sec:width}

\subsection{Method}
For bit-flip noise, $\eps=\mathbb E_\sigma|A_-(\sigma)/A_+(\sigma)|$, where $A_\pm(\sigma)=\sum_{e:\,\mathrm{syn}(e)=\sigma}\Pr(e)\,(\pm1)^{\pi(e)}$ and $\pi(e)$ is the parity of flips on row $0$. Both sums are traces of products of $d$ column transfer matrices of size $2^{w-1}$ (the state is the column of vertical-edge flips; given two adjacent columns and the syndrome, the horizontal flips are fixed up to one bit). We evaluate $A_-/A_+$ exactly for each sampled syndrome and average over $3000$ syndromes per point. For small strips the sum over all syndromes is done exactly. The transfer matrix agrees with the Walsh--Hadamard evaluation of Eq.~\eqref{eq:identity} to $10^{-8}$, and the latter agrees with brute-force reduced density matrices of the full $18$-qubit state (for $d=3$, regions up to $11$ edges, all three channels) to $10^{-15}$.

\subsection{Exponential decay in $d$}
At fixed $w$ and $p$, $\ln\eps$ is linear in $d$ to within statistical error (Fig.~\ref{fig:main}a), defining a rate
\begin{equation}
\varphi(w,p)=-\lim_{d\to\infty}\tfrac1d\ln\eps(d,w,p).
\end{equation}
Physically, $\varphi$ is the density per unit length of uncorrectable crossings of the strip. Table~\ref{tab:phi} lists $\varphi$ from fits over $d=8,\dots,32$.

\begin{table}[t]\centering\small
\caption{Decay rate $\varphi(w,p)=-\partial\ln\eps/\partial d$ for bit-flip noise, from fits over $d=8,16,24,32$ (2000 syndrome samples per point). Rows $w=1,2$ are the exact value $-\ln(1-2p)$, which both share (Corollary~\ref{cor:tworow}); the transfer-matrix estimates for $w=2$ agree with it to within $4\%$, the size of the Monte Carlo error where $\eps$ is smallest.}\label{tab:phi}
\begin{tabular}{cccccccc}\toprule
$w$ & $p=0.02$ & $p=0.04$ & $p=0.06$ & $p=0.08$ & $p=0.10$ & $p=0.12$ & $p=0.14$\\\midrule
1 & 0.0408 & 0.0834 & 0.1278 & 0.1744 & 0.2231 & 0.2744 & 0.3285\\
2 & 0.0408 & 0.0834 & 0.1278 & 0.1744 & 0.2231 & 0.2744 & 0.3285\\
3 & 0.0070 & 0.0255 & 0.0526 & 0.0852 & 0.1294 & 0.1745 & 0.2347\\
4 & 0.0028 & 0.0140 & 0.0327 & 0.0609 & 0.1008 & 0.1420 & 0.1950\\
5 & 0.0009 & 0.0059 & 0.0205 & 0.0432 & 0.0751 & 0.1207 & 0.1751\\
6 & 0.0004 & 0.0037 & 0.0132 & 0.0346 & 0.0647 & 0.1065 & 0.1578\\
7 & 0.0001 & 0.0019 & 0.0098 & 0.0264 & 0.0552 & 0.0939 & 0.1450\\
\bottomrule\end{tabular}\end{table}

\subsection{Three regimes}
\begin{figure}[t]\centering
\includegraphics[width=\textwidth]{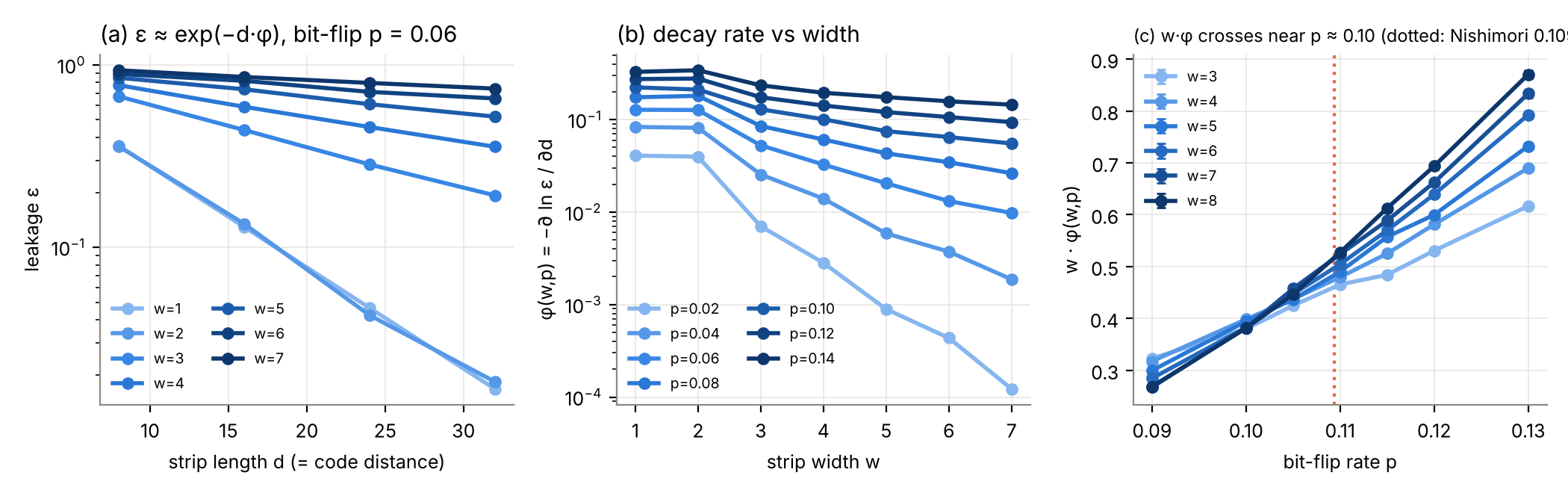}
\caption{Strip leakage under bit-flip noise. (a) $\eps(d,w,p=0.06)$ versus $d$: straight lines on a log scale; $w=1$ and $w=2$ coincide. (b) Decay rate $\varphi(w,p)$ versus width, $p=0.02,\dots,0.14$. (c) Width-scaled rate $w\varphi$ for $w=3,\dots,8$ ($d=8,\dots,40$, 3000 samples per point); dotted line: Nishimori point.}\label{fig:main}
\end{figure}

Below threshold, $\varphi(w,p)$ decreases roughly exponentially in $w$ (Fig.~\ref{fig:main}b), with an even--odd structure inherited from the distance $\lceil w/2\rceil$ of Proposition~\ref{prop:smallp}. Writing $\varphi\simeq a(p)e^{-w/\xi(p)}$, the leakage is close to one once $d\,\varphi\ll1$, i.e.
\begin{equation}
w\gtrsim w^*(d,p)=\xi(p)\ln\!\big(a(p)\,d\big).
\label{eq:wstar}
\end{equation}

\begin{figure}[t]\centering
\includegraphics[width=0.5\textwidth]{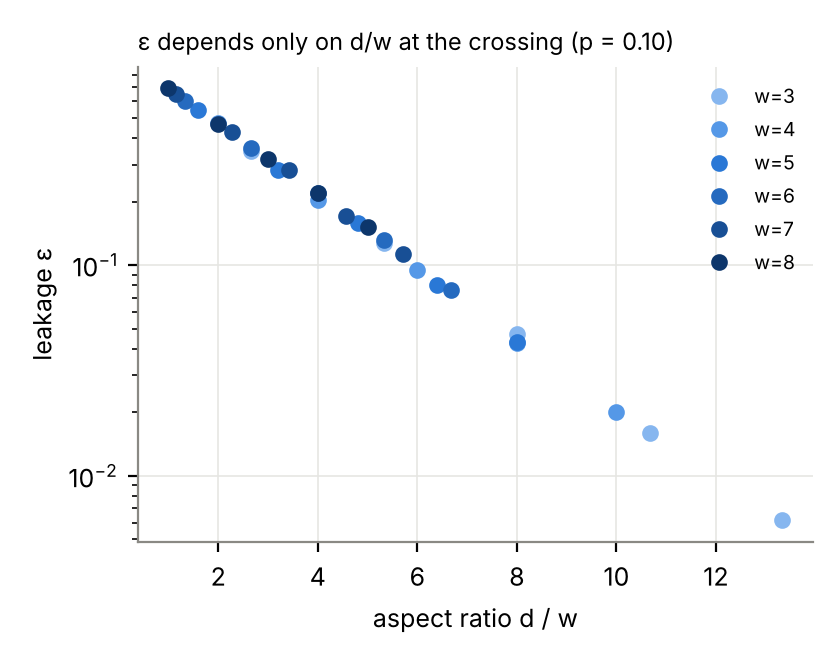}
\caption{At $p=0.10$ the leakage for $w=3,\dots,8$ and $d=8,\dots,40$ falls on a single curve of the aspect ratio $d/w$; the collapse is tight for $d/w\lesssim8$ and noisier beyond, where $\eps$ is small.}\label{fig:collapse}
\end{figure}

At a critical point of the underlying disordered Ising model, the free-energy cost per unit length of a twist in a strip of width $w$ scales as $1/w$ by finite-size scaling, and one expects $w\varphi$ to be independent of $w$ \cite{Honecker2001}. Fig.~\ref{fig:main}c shows $w\varphi(w,p)$ for $w=3,\dots,8$: the curves decrease with $w$ for $p$ below the crossing and increase above it, and cross at $p_\times\approx0.10$ (Table~\ref{tab:cross}). At the crossing, $\eps$ depends on $d$ and $w$ only through $d/w$ (Fig.~\ref{fig:collapse}). At $p=0.10$ the scaled rate is $w\varphi=0.38$--$0.40$ for all $w=3,\dots,8$.

For $w=4,\dots,8$ the pairwise crossings lie between $0.100$ and $0.106$ (Table~\ref{tab:cross}; we exclude pairs involving $w=3$, whose even--odd mismatch with $w=4$ shifts the crossing). These widths are too small to locate the threshold, so we extend them below.

\subsection{Wider strips: quenched and annealed rates}\label{sec:wide}
The column transfer matrix is an XOR convolution: $(vT_\pm)[b]=r^{|b|}\sum_av[a]f_\pm(\sigma\oplus a\oplus b)$ with $f_\pm(t)=r^{k(t)}\pm r^{w-k(t)}$ and $r=p/(1-p)$, so a fast Walsh--Hadamard transform applies it in $O(w\,2^w)$ operations instead of $O(4^w)$. This makes $w=16$ cheap. On an open strip of length $L$ (first vertical column weighted by its prior, last summed freely), $\rho=A_-/A_+$ is exactly the posterior expectation of $(-1)^\pi$ given the syndrome. We validated the fast step against the dense matrices to $10^{-16}$, the posterior against brute-force enumeration of all flip patterns on small strips to $10^{-14}$, and the Nishimori identity $\mathbb E[\rho\,(-1)^{\pi}]=\mathbb E[\rho^2]$, which holds exactly for this posterior, to within $|z|\le1.4$ on ten sampled $(w,p,L)$.

Two rates follow from the same samples. The \emph{annealed} rate $\varphi_a=-\lim L^{-1}\ln\mathbb E|\rho_L|$ is the one that governs $\eps$ and that we have used so far. The \emph{quenched} rate $\varphi_q=-\lim L^{-1}\mathbb E\ln|\rho_L|$, the Lyapunov gap between the two transfer-matrix products, is the domain-wall free energy per unit length used in transfer-matrix studies of the random-bond Ising model \cite{Honecker2001}. By Jensen's inequality $\varphi_a\le\varphi_q$, which holds at every point we computed. The two rates answer different questions. The annealed rate governs the average advantage, which is the operational quantity; the quenched rate describes the typical decay across disorder realizations and is the better-conditioned estimator of where the transition lies. We therefore locate the crossing with the quenched rate, use the annealed crossings as a separate check, and explain below why the operational threshold is fixed by Theorem~\ref{thm:identity} rather than by either fit. We obtained $\varphi_q$ for $w=4,\dots,16$ and $p=0.090,\dots,0.120$ from 16 strips of length $3000$ per point (after $300$ columns of burn-in), increased to 64 strips for $w=12,\dots,16$ at $p=0.100,\dots,0.115$, and $\varphi_a$ for $w=6,\dots,12$ from $3000$ strips of lengths $24$--$72$.

\begin{figure}[t]\centering
\includegraphics[width=\textwidth]{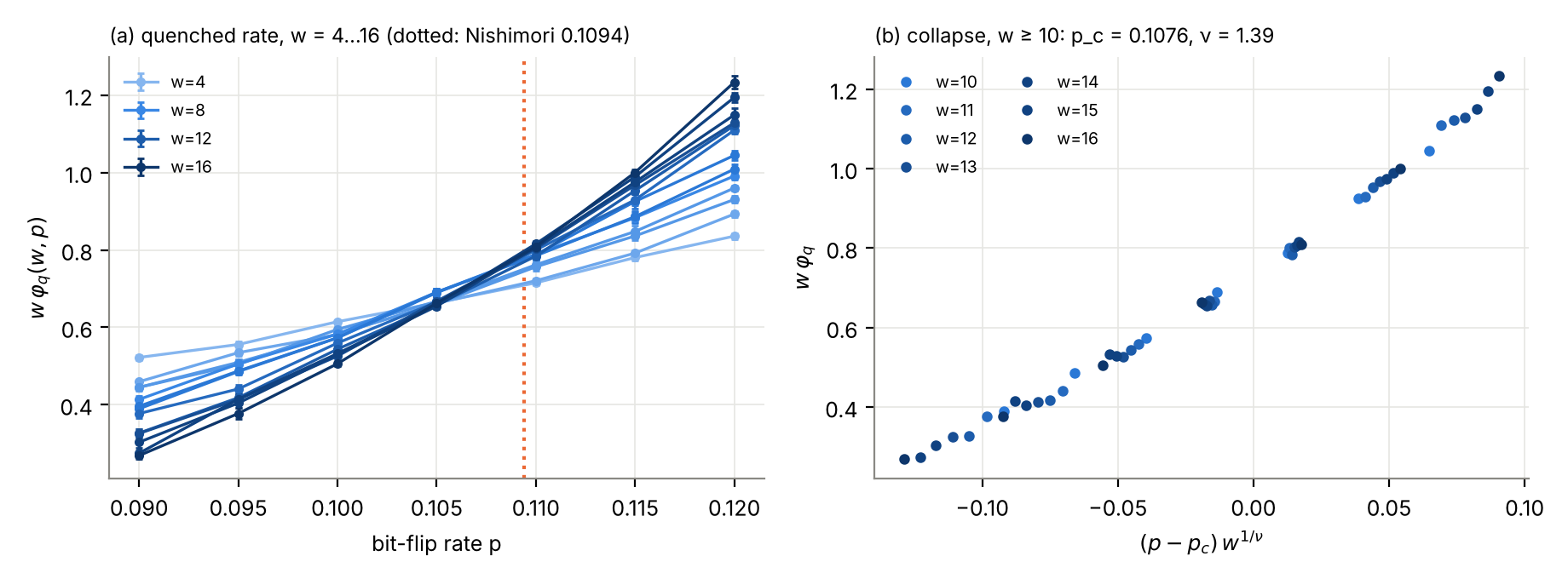}
\caption{Quenched rate from the fast transfer matrix. (a) $w\varphi_q(w,p)$ for $w=4,\dots,16$ (16 strips of length 3000 per point, 64 for $w\ge12$ at $p=0.100$--$0.115$; error bars one standard error). (b) Finite-size-scaling collapse for $w\ge10$ with $p_c=0.1076$, $\nu=1.39$.}\label{fig:quenched}
\end{figure}

Fig.~\ref{fig:quenched} shows $w\varphi_q$. A finite-size-scaling collapse $w\varphi_q=F\big((p-p_c)w^{1/\nu}\big)$ with a cubic $F$ gives $p_c=0.1076\pm0.0004$ and $\nu=1.39\pm0.07$ for $w\ge10$ (bootstrap errors; $\chi^2/\mathrm{dof}=1.2$). These statistical errors understate the uncertainty, so we repeated the fit over the choices it depends on: the smallest width included ($w\ge8$, $10$, $12$), the order of $F$ ($2$--$4$), the $p$ window ($0.090$--$0.120$ or $0.095$--$0.115$), and all, even or odd widths. Table~\ref{tab:fss} summarizes the fits that use at least three widths and have $\chi^2/\mathrm{dof}<3$. Every such fit puts the crossing between $0.105$ and $0.108$, which is $0.001$--$0.004$ below the Nishimori point $p_N=0.1094\pm0.0002$. The estimate rises from $0.1061$ ($w\ge6$) to $0.1076$ ($w\ge10$) but falls back to about $0.106$ for $w\ge12$, where only five widths remain, so the numerics do not show a monotone approach to $p_N$ at the widths reached. Our strips have open boundaries, whose surface corrections a collapse without correction-to-scaling terms does not absorb. An offset of this size is therefore not evidence against $p_N$, but the numerics alone locate the threshold only to within a few thousandths. The exponent is less well determined still: $\nu$ ranges over $1.26$--$2.03$ depending on the fit, compatible with the value $1.33\pm0.03$ reported at the Nishimori point \cite{Honecker2001} but not a test of it. The annealed rate crosses in the same region, between $0.097$ and $0.113$ for same-parity pairs at $w=6,\dots,12$, with larger scatter because $\mathbb E|\rho_L|$ is estimated from shorter strips.

\begin{table}[t]\centering\footnotesize
\caption{Robustness of the finite-size-scaling collapse of $w\varphi_q$ to fit choices: order of $F$ (2--4), $p$ window ($0.090$--$0.120$ or $0.095$--$0.115$) and widths used (all, even, odd), for each smallest width $w_{\min}$. Only fits with at least three widths and $\chi^2/\mathrm{dof}<3$ are included. The Nishimori point is $p_N=0.1094\pm0.0002$ with $\nu=1.33\pm0.03$ \cite{Honecker2001}.}\label{tab:fss}
\begin{tabular}{cccccc}\toprule $w_{\min}$ & fits & $p_c$ range & median $p_c$ & $\nu$ range & median $\nu$\\\midrule
8 & 18/18 & 0.1068--0.1078 & 0.1072 & 1.32--1.44 & 1.38\\
10 & 18/18 & 0.1053--0.1080 & 0.1075 & 1.26--2.03 & 1.41\\
12 & 12/12 & 0.1056--0.1064 & 0.1060 & 1.53--2.01 & 1.66\\
\bottomrule\end{tabular}\end{table}

\paragraph{Which threshold is operational?} The leakage is governed by $\varphi_a$, the collapse uses $\varphi_q$, and the two are different functions ($\varphi_a<\varphi_q$ at every point we computed). Jensen's inequality orders them but does not by itself make their critical points coincide, so the quenched collapse alone does not locate the threshold of $\eps$. Theorem~\ref{thm:identity} does. For a strip of width $w=d/\alpha$, $\eps=1-2\Pml$ for a cylinder surface code of fixed aspect ratio $\alpha$ whose logical failures are error chains across its width. Below the optimal threshold of the surface code, $\Pml\to0$ as the code grows and $\eps\to1$; above it, $\Pml\to1/2$ and $\eps\to0$. Under bit-flip noise that optimal threshold is the Nishimori point \cite{Dennis2002,WangHarringtonPreskill2003}, a bulk transition of the random-bond Ising model that does not depend on the aspect ratio. The operational threshold is therefore $p_N$ by Theorem~\ref{thm:identity} and this known result. The quenched collapse and the annealed crossings are consistent with this threshold within the finite-width uncertainty described above; they support the identification but do not establish it on their own.

\begin{table}[t]\centering\small
\caption{Crossings of the annealed $w\varphi(w,p)$ between pairs of widths ($d\le40$, torus), by linear interpolation in $p$; errors from resampling the fitted $\varphi$ within their standard errors. Pairs involving $w=3$ are listed for completeness but excluded from the quoted range $0.100$--$0.106$; Sec.~\ref{sec:wide} extends the analysis to $w=16$.}\label{tab:cross}
\begin{tabular}{lccccc}\toprule
pairs, same parity & $(3,5)$ & $(4,6)$ & $(5,7)$ & $(6,8)$ &\\
$p_\times$ & $0.0963(11)$ & $0.1032(9)$ & $0.1019(7)$ & $0.1016(36)$ &\\\midrule
pairs, adjacent & $(4,5)$ & $(5,6)$ & $(6,7)$ & $(7,8)$ & $(3,4)$\\
$p_\times$ & $0.1037(34)$ & $0.1027(13)$ & $0.1001(23)$ & $0.1059(59)$ & $0.0925(16)$\\\bottomrule
\end{tabular}\end{table}


\subsection{What this means for an eavesdropper}
The three regimes answer the question of the title.
\begin{itemize}
\item \emph{Below threshold} ($p<p_N$), a strip of width $w^*\sim\xi(p)\ln d$ learns the bit with advantage close to one. The width an eavesdropper needs grows only logarithmically with the code distance, even though the noiseless code distance against local regions is $d$. Conversely, a strip of fixed width learns an amount that decays as $e^{-d\varphi}$.
\item \emph{At threshold}, the leakage is a function of the aspect ratio $d/w$ alone: a strip must cover a finite fraction of the torus to learn a finite amount.
\item \emph{Above threshold}, $w\varphi$ grows with $w$, so at any fixed aspect ratio $d/w$ the leakage vanishes as the system grows. We expect $\varphi$ to saturate at a finite value deep in this regime, but at $p=0.13$ it is still decreasing slowly at $w=8$, so we do not claim saturation. In this regime the bit is also unrecoverable by a global decoder, so the hiding is not useful as a memory.
\end{itemize}

\section{Design rules for untrusted readout}\label{sec:design}

The results above translate into three rules for anyone who must hand part of a code's readout to a party they do not fully trust, whether a decoder module or hardware that shares a readout line.

\paragraph{Rule 1: geometry, not noise, provides the protection.} Theorem~\ref{thm:identity} gives $\eps_A=0$ at every noise strength whenever $A$ supports no representative of the logical. A simple sufficient condition is the following.

\begin{proposition}[One missing column]\label{prop:column}
On the torus, if $A$ contains none of the horizontal edges $h(i,j)$, $0\le i<d$, for some column $j$, then $\eps_A=0$ for every Pauli channel. For the rotated planar code read out in the $Z$ basis, if $A$ contains no data qubit of some column, then $\eps_A=0$.
\end{proposition}
\begin{proof}
$X_0$ multiplied by stars is a column of $X$ on the horizontal edges of any chosen column $j$. Every representative of $Z_0$ anticommutes with every representative of $X_0$, so it acts on at least one edge of every such column. A region that misses a whole column therefore supports no representative of $Z_0$, and Theorem~\ref{thm:identity} gives $\eps_A=0$. The planar argument is identical, with $X_L$ moved to any column by $X$ checks. We confirmed the planar statement by exact evaluation on random regions missing a column for $d=3,5$, including the whole code minus one column.
\end{proof}

A single missing column is enough: a region may hold all but one column of the code and still learn nothing. Conversely, once a region does span the code, noise is a weak defence. Table~\ref{tab:dhalf} gives the code distance at which a spanning strip's advantage has fallen only to $1/2$. At a flip rate of $10^{-3}$ a single spanning row still holds half the advantage at $d\approx350$, a three-row strip at $d\approx4\times10^4$, and a five-row strip only at $d$ of several million. Lower physical error rates make this worse, not better, since the protection of a spanning region comes only from noise.

\begin{table}[t]\centering\footnotesize
\caption{Code distance $d_{1/2}=\ln2/\varphi(w,p)$ at which a strip of width $w$ that spans the code still gives an eavesdropper advantage $1/2$, under bit-flip rate $p$. Row $w=1$ exact; other values from Monte Carlo transfer-matrix fits of $\ln\eps$ over $d=16,32,48$; $^\dagger$ from the small-$p$ form $\varphi\simeq a_wp^{\lceil w/2\rceil}$ with $a_1=2$ (exact) and $a_3\approx19$, $a_5\approx128$ fitted where $\varphi$ is resolved and extrapolated to $p\to0$, so order-of-magnitude only.}\label{tab:dhalf}
\begin{tabular}{cccccc}\toprule $w$ & $p=0.001$ & $p=0.002$ & $p=0.005$ & $p=0.01$ & $p=0.02$\\\midrule
1 & $346$ & $173$ & $69$ & $34$ & $17$\\
3 & $3.6\times10^{4}{}^\dagger$ & $9.2\times10^{3}$ & $1.5\times10^{3}$ & $389$ & $108$\\
5 & $5.4\times10^{6}{}^\dagger$ & $6.8\times10^{5}{}^\dagger$ & $4.3\times10^{4}{}^\dagger$ & $5.5\times10^{3}$ & $788$\\
\bottomrule\end{tabular}\end{table}

\paragraph{Collusion and repeated exposure.} Theorem~\ref{thm:identity} and Proposition~\ref{prop:column} concern one observer of one region during one exposure of the logical. Observers who pool their outcomes are a single observer of the union of their regions, so the guarantee must be applied to that union: two regions that each miss a column can together span the code and learn the bit outright. Proposition~\ref{prop:column} holds for any number of colluders as long as their union still misses a column. Repeated exposure behaves the same way. If the same logical value is exposed $k$ times with independent noise, the observer holds $k$ copies of $\rho_A(b)$. When $A$ supports no representative, $\rho_A(0)=\rho_A(1)$ and every number of copies gives exactly zero. When it does, the advantage grows with $k$: because the two reduced states commute (Theorem~\ref{thm:identity}), a coupling argument gives $\eps\le\eps^{(k)}\le1-(1-\eps)^k$, and $\eps^{(k)}\to1$ for any $\eps>0$. Lattice-surgery merges repeat the seam measurement for about $d$ rounds; those copies share data-qubit errors and are therefore not independent, and we do not compute their leakage. The noise-based protection of Table~\ref{tab:dhalf} is a single-exposure, non-colluding figure; the geometric guarantee is not.

\paragraph{Rule 2: decoder windows and readout groupings must not span.} For partitioned decoding \cite{Lin2024}, the rule applies to the data that carry logical outcomes, not to ordinary syndrome rounds: during a transversal readout or a lattice-surgery merge, each module's region, including its buffer, should miss at least one column of the logical being read out (and one row, for the conjugate basis), and so should the union of any modules that could share data, and the parity that defines the logical outcome should be assembled only on trusted hardware. For multiplexed readout, the data qubits sharing a feedline, together with those on any feedline the same party can observe, should miss at least one column for $Z$-basis readout and at least one row for $X$-basis readout; compact blocks smaller than the code in both directions satisfy both, whereas a feedline along a full row or column does not. Table~\ref{tab:side} shows what is at stake. A side channel that reads each qubit correctly with probability $a=0.99$ and covers one row of a $d=7$ code leaks the logical readout with advantage $0.87$; covering three rows it leaks $0.99$. Even an $80\%$-accurate channel covering three rows gives $0.09$ at $d=7$. Missing one column, it gives exactly $0$.

\begin{table}[t]\centering\footnotesize
\caption{Logical leakage through a per-qubit side channel of accuracy $a$ (each data-qubit outcome inferred correctly with probability $a$, independently), for a channel covering the top $w$ rows of a distance-$d$ rotated surface code; exact evaluation of Eq.~\eqref{eq:identity} with flip rate $1-a$ and no other noise. A channel that misses at least one entire column of data qubits leaks exactly $0$ at any $a$ (Proposition~\ref{prop:column}).}\label{tab:side}
\begin{tabular}{ccccc}\toprule $a$ & $d$ & one row & two rows & three rows\\\midrule
0.8 & 3 & 0.216 & 0.216 & 0.383\\
0.8 & 5 & 0.078 & 0.078 & 0.186\\
0.8 & 7 & 0.028 & 0.028 & 0.091\\
\midrule
0.9 & 3 & 0.512 & 0.512 & 0.761\\
0.9 & 5 & 0.328 & 0.328 & 0.609\\
0.9 & 7 & 0.210 & 0.210 & 0.488\\
\midrule
0.95 & 3 & 0.729 & 0.729 & 0.926\\
0.95 & 5 & 0.590 & 0.590 & 0.868\\
0.95 & 7 & 0.478 & 0.478 & 0.814\\
\midrule
0.99 & 3 & 0.941 & 0.941 & 0.997\\
0.99 & 5 & 0.904 & 0.904 & 0.994\\
0.99 & 7 & 0.868 & 0.868 & 0.990\\
\midrule
0.999 & 3 & 0.994 & 0.994 & 1.000\\
0.999 & 5 & 0.990 & 0.990 & 1.000\\
0.999 & 7 & 0.986 & 0.986 & 1.000\\
\bottomrule\end{tabular}\end{table}

\paragraph{Rule 3: a second row is free under uniform noise, but not on real hardware.} By Corollary~\ref{cor:tworow}, a region that already spans the code with one row learns nothing more from a second row under i.i.d.\ noise, and Table~\ref{tab:side} shows the same for side channels with uniform accuracy. The hardware data of Sec.~\ref{sec:hw} show that this does not survive non-uniform noise: an adversary who learns which qubits are cleaner gains from the second row, and a model of independent flips with per-qubit rates accounts for $90$--$99\%$ of that gain. Any partition that cannot avoid spanning should therefore be assessed with the device's actual noise, measured rather than inferred: Sec.~\ref{sec:hw} shows that calibration data rank rows reliably only on small circuits and underestimate native noise by up to $42\%$.

\section{Depolarizing noise: correlations help the eavesdropper}\label{sec:depol}
For $w\le2$ the strip supports no stars and Corollary~\ref{cor:tworow} applies to every Pauli channel. For $w\ge3$ the stars in the interior of the strip detect $Z$ components of errors, which under depolarizing noise are correlated with flips through $Y$ errors. Table~\ref{tab:depol} compares the exact depolarizing leakage with a bit-flip channel of the same flip probability $q=2p/3$. The correlations raise the leakage by $1$--$8\%$ in the range studied, with the gain growing with $p$ and $d$. This is the local analogue of the higher optimal threshold of the toric code under depolarizing noise.

\begin{table}[t]\centering\small
\caption{Exact strip leakage under depolarizing noise versus bit-flip noise with the same flip probability $q=2p/3$ (Walsh--Hadamard evaluation of Eq.~\eqref{eq:identity}). $r$ is the rank of $S_A$.}\label{tab:depol}
\begin{tabular}{cccccccc}\toprule $d$ & $w$ & $r$ & $p$ & $\eps_{\mathrm{depol}}$ & $\eps_{\mathrm{flip}}(2p/3)$ & ratio\\\midrule
4 & 3 & 12 & 0.06 & 0.9251 & 0.9079 & 1.019\\
4 & 3 & 12 & 0.12 & 0.7412 & 0.7131 & 1.039\\
5 & 3 & 15 & 0.06 & 0.9067 & 0.8845 & 1.025\\
5 & 3 & 15 & 0.12 & 0.6838 & 0.6495 & 1.053\\
6 & 3 & 18 & 0.06 & 0.8889 & 0.8630 & 1.030\\
6 & 3 & 18 & 0.12 & 0.6323 & 0.5949 & 1.063\\
7 & 3 & 21 & 0.06 & 0.8715 & 0.8420 & 1.035\\
7 & 3 & 21 & 0.12 & 0.5852 & 0.5451 & 1.074\\
5 & 4 & 25 & 0.06 & 0.9593 & 0.9361 & 1.025\\
5 & 4 & 25 & 0.12 & 0.7909 & 0.7417 & 1.066\\
\bottomrule\end{tabular}\end{table}

\section{A strip adversary emulated on superconducting hardware}\label{sec:hw}

The hardware experiment asks something the uniform-noise model cannot: whether differences in the reliability of individual qubits change what a regional observer can learn. We compare a model-free attacker trained on the data with maximum-likelihood decoders that assume uniform and per-qubit error rates, which separates the effect of holding more qubits from the effect of exploiting non-uniform hardware. The adversary is emulated offline from the full readout; the aim is to test the predicted leakage structure under real circuit noise, not to demonstrate a live side-channel attack (see the end of this section).

\paragraph{Protocol.} We prepared $|0_L\rangle$ and $|1_L\rangle$ of the rotated surface code with $d=3$ and $d=5$ on two IBM Heron r2 processors, \texttt{ibm\_marrakesh} and \texttt{ibm\_kingston}, in five runs over three days (3--5 November 2026): Marrakesh on each of the three days and Kingston on the second and third. Each run used $20{,}000$ shots per logical state and distance in a single four-circuit job ($23$--$24$\,s of QPU time per job, $118$\,s in total; job identifiers in Appendix~\ref{app:jobs}). The encoder is measurement-free: the $X$ checks are ordered by peeling so that each owns a pivot qubit absent from all earlier checks, and each is applied as a Hadamard on its pivot followed by CNOTs to its other qubits ($8$ and $28$ CNOTs before routing; $10$ and $100$ CZ gates after routing on the heavy-hex lattice, in every run). The $|1_L\rangle$ circuit is the transpiled $|0_L\rangle$ circuit with $X$ gates inserted before readout on the physical qubits carrying $X_L$, so the two differ only in $b$. All data qubits, pivots included, start in $|0\rangle$, which satisfies every $Z$ check and has $Z_L=+1$; each pivot's Hadamard and CNOT fan-out adds only $X$-check patterns, which commute with the $Z$ checks, so the $Z$ checks remain satisfied while the $X$ checks are prepared. Finally, every data qubit is measured in the $Z$ basis. For every job we saved the exact transpiled circuits and the backend calibration in force when it ran.

The rotated code has the same strip structure as the torus. For the planar strip of the top $w$ rows, the exact calculation of Eq.~\eqref{eq:identity} gives $\eps_{A_1}=\eps_{A_2}=(1-2q)^d$ and a jump at $w=3$ (e.g.\ $0.328,0.328,0.609,0.633$ for $w=1,\dots,4$ at $d=5$, $q=0.1$).

An eavesdropper holding the top $w$ rows sees only those bits. For noise acting as $X$ errors before readout, $Z$-basis readout of the strip followed by decoding is the optimal measurement of Theorem~\ref{thm:identity}, so the measured advantage $\hat\eps=2\Pr[\text{correct}]-1$ is an unbiased estimate of what this attack achieves and a lower bound on $\eps_{A_w}$ for the prepared noisy state. To sweep the noise strength we flip each recorded bit independently with probability $p_{\rm inj}$, which is exactly equivalent to an $X$ error immediately before readout. We score two attackers: the exact ML strip decoder under an i.i.d.\ model at the native flip rate estimated from $Z$-check violations, and a model-free attacker that learns a lookup table on (strip syndrome, strip logical parity) from half the shots and is evaluated on the other half. Because it is inferred from the final readout, the native flip rate $q_{\rm nat}$ lumps together every process that flips a measured bit (gate errors during encoding, including correlated two-qubit errors and errors spread by routing SWAPs, idle decay, and readout errors); $Z$-type errors are invisible to it, and it is an effective uniform rate fitted to noise that is neither uniform nor independent. Where the learned attacker is evaluated with injected noise, we average over eight independent injection and train/test seeds.

\begin{figure}[t]\centering
\includegraphics[width=0.85\textwidth]{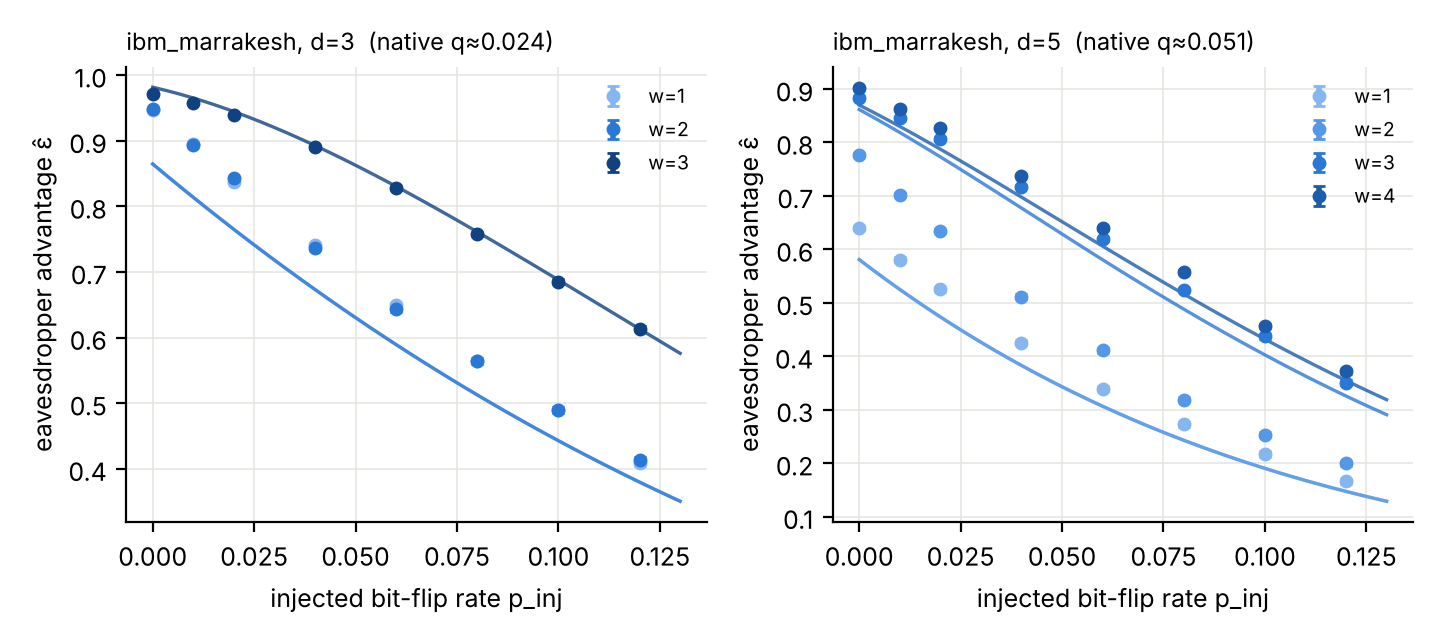}
\caption{The first hardware run (\texttt{ibm\_marrakesh}, 3 November 2026). Points: best of the model and learned attackers (40{,}000 shots per point; error bars one standard error, smaller than the markers). Lines: i.i.d.\ theory at the composed native+injected flip rate; $w=1$ and $w=2$ share a line. The other four runs are summarized in Table~\ref{tab:hw}.}\label{fig:hw}
\end{figure}

\paragraph{Results.} Table~\ref{tab:hw} summarizes the five runs, Table~\ref{tab:quality} the quality of the prepared states, and Fig.~\ref{fig:hw} shows the first run in full.
\begin{enumerate}
\item \emph{The first gain is at $w=3$, on every run.} The three-row strip beats the single row in all ten run--distance combinations, by $0.024$--$0.111$ at $d=3$ and $0.160$--$0.347$ at $d=5$. This is a consistency check rather than a sharp test, since the planar formula predicts it.
\item \emph{Wider strips follow the i.i.d.\ theory where noise is uniform.} For $w\ge3$ and $p_{\rm inj}\ge0.02$ the learned advantage matches the i.i.d.\ prediction at the composed rate to within $0.01$ on average for both Kingston runs at both distances. On Marrakesh it exceeds the prediction by $0.00$--$0.03$ at $d=3$ and $0.03$--$0.09$ at $d=5$; it never falls below it by more than $0.007$. An attacker on a non-uniform device does better than a uniform model predicts, never worse.
\item \emph{The second row helps only through non-uniform, essentially independent noise.} Under i.i.d.\ noise the optimal two-row decoder never uses the second row (Corollary~\ref{cor:tworow}); the model decoder therefore returns $\hat\eps_{w=2}=\hat\eps_{w=1}$ by construction, and that equality is not a test. On the planar code a sharper statement holds for independent flips of arbitrary, non-identical rates: every representative of $Z_L$ supported in the two-row strip has exactly one qubit in each column, since multiplying the top row by a $Z$ check only exchanges which row is used in the columns that check covers. Adding i.i.d.\ flips at rate $p_{\rm inj}$ therefore multiplies every term of Eq.~\eqref{eq:identity}, for both $w=1$ and $w=2$, by $(1-2p_{\rm inj})^d$, so the relative gain $\hat\eps_{w=2}/\hat\eps_{w=1}-1$ is exactly invariant (we confirmed this numerically to all digits for random rates). A shrinking absolute gain under injected noise is therefore no evidence about the origin of the gain; the relative gain is the quantity to test. Table~\ref{tab:tworow} shows that it is invariant within seed spread on all five runs, and that a per-qubit-rate ML decoder, with flip rates fitted on half the shots, reproduces $90$--$99\%$ of the learned attacker's gain on the other half. Independent but non-uniform flips thus account for almost all of the effect; the residual, at most $0.011$, bounds what correlations add. At $d=3$ both attackers find no gain, because the fitted rates make the top row the cleaner of the two in every column.
\item \emph{The gain depends on how the top two rows compare.} The largest two-row gain ($0.270$, Marrakesh on 5 November) coincides with a single-row advantage of only $0.467$, caused by poor qubits in the top row; the second row lets the attacker route around them, column by column. The smallest ($0.061$, Kingston on 5 November) comes with top rows of similar quality. Run to run, native flip rates varied from $2.0\%$ to $4.2\%$ at $d=3$ and from $3.6\%$ to $8.2\%$ at $d=5$, with Kingston the cleaner device on both days it was used.
\item \emph{Larger codes hide more from a fixed-width strip}, although the comparison is confounded: at $w=3$ the mean advantage over runs drops from $0.973$ at $d=3$ to $0.870$ at $d=5$, but the deeper $d=5$ circuit also has twice the native flip rate.
\end{enumerate}

\begin{table}[t]\centering\footnotesize\setlength{\tabcolsep}{4pt}
\caption{Five hardware runs ($20{,}000$ shots per state per distance; $4$ circuits and $23$--$24$\,s of QPU time each, $118$\,s in total). Native flip rate from $Z$-check violations; learned-attacker advantage with no injected noise (standard error $\le0.004$); and the learned two-row gain $\hat\eps_{w=2}-\hat\eps_{w=1}$ at $d=5$ for $p_{\rm inj}=0,0.06,0.12$ (mean over 8 injection and train/test seeds; seed spread $\le0.008$).}\label{tab:hw}
\begin{tabular}{lccccccccc}\toprule
 & \multicolumn{4}{c}{$d=3$} & \multicolumn{5}{c}{$d=5$}\\\cmidrule(lr){2-5}\cmidrule(lr){6-10}
run & $q_{\rm nat}$ & $w=1$ & $w=2$ & $w=3$ & $q_{\rm nat}$ & $w=1$ & $w=3$ & $w=4$ & two-row gain\\\midrule
Marrakesh, 3 Nov & 2.4\% & 0.947 & 0.948 & 0.971 & 5.1\% & 0.641 & 0.884 & 0.901 & 0.135/0.067/0.030\\
Marrakesh, 4 Nov & 4.2\% & 0.850 & 0.860 & 0.961 & 8.2\% & 0.662 & 0.829 & 0.848 & 0.080/0.042/0.014\\
Kingston, 4 Nov & 2.2\% & 0.915 & 0.915 & 0.982 & 3.9\% & 0.749 & 0.908 & 0.934 & 0.099/0.053/0.024\\
Marrakesh, 5 Nov & 3.0\% & 0.942 & 0.941 & 0.968 & 7.0\% & 0.467 & 0.813 & 0.839 & 0.270/0.136/0.064\\
Kingston, 5 Nov & 2.0\% & 0.888 & 0.887 & 0.984 & 3.6\% & 0.753 & 0.916 & 0.934 & 0.061/0.022/0.009\\
\bottomrule\end{tabular}\end{table}

\begin{table}[t]\centering\footnotesize
\caption{Two-row test at $d=5$. Relative gain $\hat\eps_{w=2}/\hat\eps_{w=1}-1$ of the learned attacker at three injected rates (mean over 8 seeds); for independent noise of any uniformity it is exactly invariant under injected i.i.d.\ noise (Sec.~\ref{sec:hw}). Absolute gain with no injected noise for the per-qubit-rate ML decoder (flip rates of the strip's qubits fitted by maximum likelihood on one half of the shots, decoding the other half) and for the learned lookup-table attacker. At $d=3$ every entry is zero within $0.010$.}\label{tab:tworow}
\begin{tabular}{lccccc}\toprule & \multicolumn{3}{c}{relative gain, learned} & \multicolumn{2}{c}{absolute gain, $p_{\rm inj}=0$}\\\cmidrule(lr){2-4}\cmidrule(lr){5-6}
run & $p_{\rm inj}=0$ & $0.06$ & $0.12$ & per-qubit ML & learned\\\midrule
Marrakesh, 3 Nov & 21.1\% & 19.8\% & 18.4\% & 0.124 & 0.135\\
Marrakesh, 4 Nov & 12.1\% & 11.9\% & 8.5\% & 0.079 & 0.080\\
Kingston, 4 Nov & 13.3\% & 13.4\% & 12.5\% & 0.099 & 0.100\\
Marrakesh, 5 Nov & 57.8\% & 54.9\% & 54.0\% & 0.268 & 0.271\\
Kingston, 5 Nov & 8.0\% & 5.4\% & 4.9\% & 0.055 & 0.061\\
\bottomrule\end{tabular}\end{table}

\begin{table}[t]\centering\footnotesize
\caption{Quality of the prepared states. Mean (range) of the $Z$-check violation rates in the final readout, the native flip rate $q_{\rm nat}$ inferred from them, and the logical error rate when the whole code is decoded with minimum-weight perfect matching.}\label{tab:quality}
\begin{tabular}{lcccccc}\toprule & \multicolumn{3}{c}{$d=3$} & \multicolumn{3}{c}{$d=5$}\\\cmidrule(lr){2-4}\cmidrule(lr){5-7}
run & check violations & $q_{\rm nat}$ & logical error & check violations & $q_{\rm nat}$ & logical error\\\midrule
Marrakesh, 3 Nov & 0.067 (0.033--0.112) & 2.4\% & 1.46\% & 0.150 (0.056--0.221) & 5.1\% & 3.31\%\\
Marrakesh, 4 Nov & 0.115 (0.068--0.159) & 4.2\% & 1.93\% & 0.220 (0.069--0.415) & 8.2\% & 6.24\%\\
Kingston, 4 Nov & 0.063 (0.029--0.084) & 2.2\% & 0.91\% & 0.118 (0.055--0.204) & 3.9\% & 2.40\%\\
Marrakesh, 5 Nov & 0.085 (0.034--0.116) & 3.0\% & 1.59\% & 0.193 (0.042--0.430) & 7.0\% & 6.40\%\\
Kingston, 5 Nov & 0.057 (0.030--0.091) & 2.0\% & 0.77\% & 0.109 (0.039--0.165) & 3.6\% & 2.17\%\\
\bottomrule\end{tabular}\end{table}

\paragraph{Predicting the hardware from its calibration.} Simulating each job's exact physical circuits with a noise model built only from that job's calibration (gate, readout and $T_1/T_2$ errors, including idle periods) underestimated the native flip rate in all ten cases, by $6$--$42\%$, ranked the rows reliably at $d=3$ but not at $d=5$, and predicted $w\ge2$ strip advantages to within $0.005$--$0.048$ (Appendix~\ref{app:jobs}, Table~\ref{tab:calpred}). Calibration data are therefore an unreliable guide to which qubits an eavesdropper should prefer on routed circuits.

\paragraph{What the experiment does and does not show.} The adversary is emulated offline: we read out every data qubit and hand the bits of the strip to the attacker in software. The experiment is therefore a hardware demonstration of regional readout leakage, with real device noise in the prepared state; it does not show that an attacker can obtain those bits through an actual shared feedline or a compromised decoder module, and a physical side channel would add its own inference error on top (Table~\ref{tab:side} shows the effect of such an error). Within that scope, the emulated attacker receives exactly what the adversary of Sec.~\ref{sec:threat} receives, the $Z$-basis readout of a region, and decodes classically. Once the encoded state is measured, the analysis is classical post-processing of $Z$-basis bits; that is precisely the optimal attack the theory identifies for flip-type noise, so the experiment tests the regional leakage--decoding identity and its strip predictions under real, correlated and non-uniform device noise. It is not a test of quantum advantage, and it does not probe the depolarizing correlation effect of Sec.~\ref{sec:depol}, which would require the eavesdropper to measure star operators inside the strip. Routing on the heavy-hex lattice roughly doubles the native flip rate from $d=3$ to $d=5$; at $d=7$ the encoder routes to about $315$ CZ gates, so we did not extend the runs to $d=7$, where native noise would swamp the distance dependence.

\section{Discussion}
\paragraph{Relation to prior work.} The noiseless access structure is the cleaning lemma and the BPT tradeoff \cite{BravyiTerhal2009,BPT2010}; the role of region size in storing quantum information is bounded in general in Ref.~\cite{Flammia2017}. Global diagnostics of the decohered toric code (coherent information, R\'enyi correlators, Markov length) locate the same threshold \cite{Bao2023,Fan2023,SangHsieh2024}; they quantify how much of the logical information survives in the whole system or can be recovered locally, whereas $\eps_A$ quantifies how much of it a single region holds. Mixed-state topological orders in two dimensions, including those of the decohered toric code, have been classified up to two-way connection by local channels \cite{EllisonCheng2025}; that classification concerns which phase a decohered state is in, not how much a given region learns about its logical content. For quantum LDPC codes under local Pauli noise, Yang \emph{et al.} show that the conditional mutual information across a buffer decays exponentially below threshold \cite{Yang2026}; that is a statement about correlations between regions separated by a buffer, whereas $\eps_A$ is the distinguishability of the two logical states on one region. We have not found earlier work that states the identity of Theorem~\ref{thm:identity} or studies strip leakage. Decoding a region on its own is decoding with its complement erased, so Theorem~\ref{thm:identity} is a statement about maximum-likelihood decoding under loss \cite{Stace2009} specialized to a chosen erasure pattern. The rate $\varphi(w,p)$ is likewise the known logical-error scaling of a long, narrow code, a failure probability per unit length set by an entropic factor and a cost that grows with the short side \cite{Dennis2002,Beverland2019}, here measured in the units of an eavesdropper's advantage. Quasi-local decoding with a buffer region has recently been studied operationally for MWPM \cite{Dehghani2026}; our strip is the complementary geometry, in which the region carries the logical itself and the decoder is optimal. The free-energy crossing that locates the Nishimori point in transfer-matrix studies of the random-bond Ising model \cite{Honecker2001} is the stat-mech counterpart of our $w\varphi$ crossing; we compute that quenched quantity alongside the annealed one that defines $\eps$ (Sec.~\ref{sec:wide}). The identification of the operational threshold with $p_N$ follows from Theorem~\ref{thm:identity} and the known optimal threshold, and the two rates are numerical evidence for it at accessible widths, not an independent derivation.

\paragraph{Limitations.} (i) The operational leakage $\eps$ is computed for $w\le8$ on the torus, the annealed rate for $w\le12$ and the quenched rate for $w\le16$ on open strips. The fitted crossing lies $0.001$--$0.004$ below $p_N$ without a monotone approach at $w\le16$, and $\nu$ is not determined at these widths; the identification of the threshold with $p_N$ rests on Theorem~\ref{thm:identity} and the known optimal threshold. (ii) The transfer-matrix method covers channels with independent $X$ and $Z$ components; depolarizing results rely on the exact Walsh--Hadamard evaluation and are limited to $|S_A|\le2^{25}$. (iii) The theory treats a single round of Pauli noise, of arbitrary correlation for Theorem~\ref{thm:identity} and i.i.d.\ for the numerics, followed by perfect measurement; circuit-level noise and noisy syndrome extraction are not modelled, although the hardware data contain both. The observer's measurement is the optimal one, and a practical eavesdropper using a suboptimal decoder learns less. (iv) The quantitative results concern one region and one exposure of the logical. Sec.~\ref{sec:design} extends the zero-leakage rule to colluding regions and bounds independent repeated exposures, but we do not compute leakage for correlated repeated exposures such as the $\sim d$ rounds of a lattice-surgery merge. (v) The hardware adversary is emulated offline from full readout, not obtained through a physical side channel; the runs reach only $d=5$, use flip-type ($Z$-basis) attacks, and compare against an i.i.d.\ model with a single native rate; the calibration-based model underestimates native noise by $6$--$42\%$.

\section{Conclusion}
The information that a region of a noisy toric code holds about the logical bit is exactly the information a decoder confined to that region can recover. For a strip that wraps the torus, this makes the leakage a finite-size decoding problem with a known phase diagram: below threshold an eavesdropper needs a width only logarithmic in the code distance, at threshold the leakage depends only on the aspect ratio, and above threshold the leakage at fixed aspect ratio vanishes with system size. On superconducting hardware the predicted width structure is visible directly, and the departures from it are a measure of how far real noise is from uniform. Code and data are provided as ancillary files.

\appendix
\section{Numerical checks}\label{app:checks}
\begin{enumerate}
\item Brute force: full state vector of the $d=3$ toric code ($18$ qubits), partial trace onto $A$, the channel applied qubit by qubit, trace distance from eigenvalues. Regions: row, two-row strip, two-row strip plus one and two interior stars, and two irregular regions; channels: depolarizing, bit-flip, dephasing; $p\in\{0,0.07,0.2\}$. Maximum deviation from Eq.~\eqref{eq:identity}: $1.0\times10^{-15}$.
\item Transfer matrix with exact summation over all syndromes versus Eq.~\eqref{eq:identity}: $(d,w)\in\{(3,2),(4,2),(4,3),(5,3),(6,3)\}$, $p\in\{0.05,0.1\}$, agreement to $10^{-8}$.
\item Monte Carlo transfer matrix versus exact values: agreement within two standard errors.
\item Corollary~\ref{cor:tworow}: $\eps_{A_2}=(1-2q)^d$ for all three channels in the brute force and for $d\le6$ in the exact evaluation.
\item Small-$p$ exponents of $1-\eps$ for $w=2,3,4,5$: $0.98$, $1.98$, $2.04$, $3.00$.
\end{enumerate}

\section{Hardware jobs}\label{app:jobs}
Table~\ref{tab:jobs} lists the five IBM Quantum jobs. Each job contained four circuits ($|0_L\rangle$, $|1_L\rangle$ at $d=3$ and $d=5$), $20{,}000$ shots each. The local time is when the shot record was saved, within a minute of execution; the calibration time is the backend's last calibration before the job. Shot records, transpiled circuits (QPY) and calibration snapshots for every job are included in the ancillary files.
\begin{table}[h]\centering\footnotesize
\caption{IBM Quantum jobs used in Sec.~\ref{sec:hw}. Times are US Eastern.}\label{tab:jobs}
\begin{tabular}{llllcc}\toprule backend & job identifier & recorded & last calibration & version & QPU time\\\midrule
\texttt{ibm\_marrakesh} & \texttt{dbl1km1kdqps7393dtq0} & 2026-11-03 12:31 & 2026-11-03 11:21 & 1.0.21 & 24\,s\\
\texttt{ibm\_marrakesh} & \texttt{dbln8r9kbkrc73co24og} & 2026-11-04 13:08 & 2026-11-04 12:51 & 1.0.21 & 24\,s\\
\texttt{ibm\_kingston} & \texttt{dbln96u74qds73dqqheg} & 2026-11-04 14:34 & 2026-11-04 14:18 & 1.0.0 & 23\,s\\
\texttt{ibm\_marrakesh} & \texttt{dbma02e74qds73drti8g} & 2026-11-05 10:26 & 2026-11-05 10:11 & 1.0.21 & 24\,s\\
\texttt{ibm\_kingston} & \texttt{dbma0f9kbkrc73cp4ik0} & 2026-11-05 10:42 & 2026-11-05 09:30 & 1.0.0 & 23\,s\\
\bottomrule\end{tabular}\end{table}

Table~\ref{tab:calpred} compares each job with a simulation of its exact transpiled circuits under a noise model built only from that job's calibration: per-pair CZ errors, per-qubit $\sqrt{X}$ errors, asymmetric readout errors, and Pauli-twirled $T_1/T_2$ decay during gates and during the idle periods of an as-late-as-possible schedule.

\begin{table}[t]\centering\footnotesize
\caption{Predicting each run from its own calibration (gate errors, readout errors, $T_1/T_2$ decay during gates and idle periods; no measured data). Fraction by which the calibration underestimates the native flip rate; rank correlation between predicted and measured row reliabilities, and whether the cleanest row is predicted correctly; mean absolute error of the learned-attacker advantage for $w\ge2$ strips at $p_{\rm inj}\in\{0,0.06\}$.}\label{tab:calpred}
\begin{tabular}{lcccccc}\toprule & \multicolumn{3}{c}{$d=3$} & \multicolumn{3}{c}{$d=5$}\\\cmidrule(lr){2-4}\cmidrule(lr){5-7}
run & noise under-est. & row rank (best?) & strip error & noise under-est. & row rank (best?) & strip error\\\midrule
marrakesh, 3 Nov & 21\% & +0.5 (yes) & 0.016 & 6\% & +0.8 (no) & 0.005\\
marrakesh, 4 Nov & 42\% & +1.0 (yes) & 0.044 & 33\% & +0.6 (yes) & 0.033\\
kingston, 4 Nov & 28\% & +1.0 (yes) & 0.007 & 11\% & +0.9 (no) & 0.012\\
marrakesh, 5 Nov & 34\% & +1.0 (yes) & 0.016 & 40\% & +0.1 (yes) & 0.048\\
kingston, 5 Nov & 27\% & +1.0 (yes) & 0.022 & 13\% & +0.0 (no) & 0.020\\
\bottomrule\end{tabular}\end{table}

\section*{Acknowledgments}
We acknowledge the use of IBM Quantum services for this work. The views expressed are those of the author, and do not reflect the official policy or position of IBM or the IBM Quantum team.

\section*{Use of AI tools}
This work was carried out with extensive assistance from Claude, an AI system developed by Anthropic. The author posed the research question and developed its initial formulation in an earlier draft, including the trace-distance leakage of a logical bit to a region, the result that local Pauli noise cannot create leakage, and the single-row law. The author also directed the development of the regional leakage--decoding identity and named it. Building on that draft and under the author's direction, Claude derived that identity and the strip results, wrote and ran the simulation and analysis code, carried out the numerical calculations, designed the hardware protocol and its analysis, searched the literature, and drafted the manuscript text. The author ran the experiments on IBM Quantum hardware, proposed and reviewed all revisions to the text, reviewed all results, and takes full responsibility for the content. Every reference was checked against its primary source, and every numerical result can be reproduced from the code and data provided as ancillary files.

\bibliographystyle{unsrt}
\bibliography{refs}

\begin{thebibliography}{10}

\bibitem{Kitaev2003}
A.~Yu. Kitaev.
\newblock Fault-tolerant quantum computation by anyons.
\newblock {\em Annals of Physics}, 303:2--30, 2003.

\bibitem{Dennis2002}
Eric Dennis, Alexei Kitaev, Andrew Landahl, and John Preskill.
\newblock Topological quantum memory.
\newblock {\em Journal of Mathematical Physics}, 43:4452--4505, 2002.

\bibitem{BPT2010}
Sergey Bravyi, David Poulin, and Barbara Terhal.
\newblock Tradeoffs for reliable quantum information storage in 2{D} systems.
\newblock {\em Physical Review Letters}, 104:050503, 2010.

\bibitem{BravyiTerhal2009}
Sergey Bravyi and Barbara Terhal.
\newblock A no-go theorem for a two-dimensional self-correcting quantum memory
  based on stabilizer codes.
\newblock {\em New Journal of Physics}, 11:043029, 2009.

\bibitem{CleveGottesmanLo1999}
Richard Cleve, Daniel Gottesman, and Hoi-Kwong Lo.
\newblock How to share a quantum secret.
\newblock {\em Physical Review Letters}, 83:648--651, 1999.

\bibitem{Lin2024}
Sophia~Fuhui Lin, Eric~C. Peterson, Krishanu Sankar, and Prasahnt Sivarajah.
\newblock Spatially parallel decoding for multi-qubit lattice surgery, 2024.
\newblock arXiv:2403.01353.

\bibitem{Skoric2023}
Luka Skoric, Dan~E. Browne, Kenton~M. Barnes, Neil~I. Gillespie, and Earl~T.
  Campbell.
\newblock Parallel window decoding enables scalable fault tolerant quantum
  computation, 2022.
\newblock arXiv:2209.08552.

\bibitem{Tan2023}
Xinyu Tan, Fang Zhang, Rui Chao, Yaoyun Shi, and Jianxin Chen.
\newblock Scalable surface code decoders with parallelization in time, 2022.
\newblock arXiv:2209.09219.

\bibitem{Maurya2024}
Satvik Maurya, Chaithanya~Naik Mude, Benjamin Lienhard, and Swamit Tannu.
\newblock Understanding side-channel vulnerabilities in superconducting qubit
  readout architectures, 2024.
\newblock arXiv:2405.08962.

\bibitem{Mehra2024}
Devika Mehra and Amir Kalev.
\newblock Towards defending crosstalk-mediated attacks in multi-tenant quantum
  computing, 2024.
\newblock arXiv:2409.14598.

\bibitem{Bao2023}
Yimu Bao, Ruihua Fan, Ashvin Vishwanath, and Ehud Altman.
\newblock Mixed-state topological order and the errorfield double formulation
  of decoherence-induced transitions, 2023.
\newblock arXiv:2301.05687.

\bibitem{Fan2023}
Ruihua Fan, Yimu Bao, Ehud Altman, and Ashvin Vishwanath.
\newblock Diagnostics of mixed-state topological order and breakdown of quantum
  memory, 2023.
\newblock arXiv:2301.05689.

\bibitem{SangHsieh2024}
Shengqi Sang and Timothy~H. Hsieh.
\newblock Stability of mixed-state quantum phases via finite {M}arkov length,
  2024.
\newblock arXiv:2404.07251.

\bibitem{Nishimori1981}
Hidetoshi Nishimori.
\newblock Internal energy, specific heat and correlation function of the
  bond-random {I}sing model.
\newblock {\em Progress of Theoretical Physics}, 66:1169--1181, 1981.

\bibitem{Honecker2001}
Andreas Honecker, Marco Picco, and Pierre Pujol.
\newblock Universality class of the {N}ishimori point in the 2{D} $\pm j$
  random-bond {I}sing model.
\newblock {\em Physical Review Letters}, 87:047201, 2001.

\bibitem{Beverland2019}
Michael~E. Beverland, Benjamin~J. Brown, Michael~J. Kastoryano, and Quentin
  Marolleau.
\newblock The role of entropy in topological quantum error correction.
\newblock {\em Journal of Statistical Mechanics: Theory and Experiment},
  2019:073404, 2019.

\bibitem{WangHarringtonPreskill2003}
Chenyang Wang, Jim Harrington, and John Preskill.
\newblock Confinement-{H}iggs transition in a disordered gauge theory and the
  accuracy threshold for quantum memory.
\newblock {\em Annals of Physics}, 303:31--58, 2003.

\bibitem{NielsenChuang2000}
Michael~A. Nielsen and Isaac~L. Chuang.
\newblock {\em Quantum Computation and Quantum Information}.
\newblock Cambridge University Press, 2000.

\bibitem{Horsman2012}
Clare Horsman, Austin~G. Fowler, Simon Devitt, and Rodney~Van Meter.
\newblock Surface code quantum computing by lattice surgery.
\newblock {\em New Journal of Physics}, 14:123011, 2012.

\bibitem{Fattal2004}
David Fattal, Toby~S. Cubitt, Yoshihisa Yamamoto, Sergey Bravyi, and Isaac~L.
  Chuang.
\newblock Entanglement in the stabilizer formalism, 2004.
\newblock arXiv:quant-ph/0406168.

\bibitem{Matsui1993}
Mitsuru Matsui.
\newblock Linear cryptanalysis method for {DES} cipher.
\newblock In {\em Advances in Cryptology -- EUROCRYPT '93}, volume 765 of {\em
  LNCS}, pages 386--397, 1994.

\bibitem{Flammia2017}
Steven~T. Flammia, Jeongwan Haah, Michael~J. Kastoryano, and Isaac~H. Kim.
\newblock Limits on the storage of quantum information in a volume of space.
\newblock {\em Quantum}, 1:4, 2017.

\bibitem{EllisonCheng2025}
Tyler~D. Ellison and Meng Cheng.
\newblock Toward a classification of mixed-state topological orders in two
  dimensions.
\newblock {\em PRX Quantum}, 6:010315, 2025.
\newblock arXiv:2405.02390.

\bibitem{Yang2026}
Tai-Hsuan Yang, Andrew~D. Kim, Akash Vijay, and Jong~Yeon Lee.
\newblock Stability of quantum {LDPC} states under local {P}auli noise, 2026.
\newblock arXiv:2610.22357.

\bibitem{Stace2009}
Thomas~M. Stace, Sean~D. Barrett, and Andrew~C. Doherty.
\newblock Thresholds for topological codes in the presence of loss.
\newblock {\em Physical Review Letters}, 102:200501, 2009.

\bibitem{Dehghani2026}
Hossein Dehghani, Sarang Gopalakrishnan, and Michael Gullans.
\newblock Error-adaptive quasi-local decoding of the toric code, 2026.
\newblock arXiv:2608.27740.

\end{thebibliography}
\end{document}